\documentclass[twocolumn,showpacs,prl,floatfix]{revtex4-1}
\usepackage{amsmath}
\usepackage{amssymb}
\usepackage{graphicx}

\newcommand{\abs}[1]{\left|#1\right|}

\newcommand{\eqRef}[1]{Eq.~(\ref{#1})}

\newcommand{\figWidth}{0.975\columnwidth}

\newcommand{\betaIceThreeSix}{1.307\:(\text{sec/cm})^{1/5}}
\newcommand{\vMaxIce}{0.262\:\text{cm/sec}}
\newcommand{\vMaxIcePadeTwoFife}{2.6\:\text{m/sec}}

\begin{document}
\title{Collision of Viscoelastic Spheres: Compact Expressions\\ for the
Coefficient of Normal Restitution}
\author{Patric M\"uller}
\author{Thorsten P\"oschel}
\affiliation
{  
Institute for Multiscale Simulation,
Universit\"at Erlangen-N\"urnberg,
N\"agelsbachstra{\ss}e 49b,
91052 Erlangen,
Germany
}

\date{\today}

\begin{abstract}
The coefficient of restitution of colliding viscoelastic spheres is 
analytically known as a complete series expansion in terms of the impact
velocity where all (infinitely many) coefficients are known. While beeing
analytically exact, this result is not suitable for applications in
efficient event-driven Molecular Dynamics (eMD) or Monte Carlo (MC) simulations.
Based on the analytic result, here we derive expressions for the coefficient of
restitution which allow for an application in efficient eMD and MC simulations
of granular Systems. 
\end{abstract}

\pacs{45.70.-n,45.50.Tn}

\maketitle

\noindent {\em Introduction and description of the system.} 
The collision of frictionless (smooth) viscoelastic spheres obeys Newton's 
equation of motion,
\begin{equation}
\label{eq:newton}
m_\text{eff}\ddot{\xi}=F\left(\dot{\xi},\xi\right),
\end{equation}
with the effective mass $m_\text{eff}\equiv m_1m_2/(m_1+m_2)$ and the 
compression $\xi\equiv R_1+R_2-\abs{\vec{r}_1-\vec{r}_2}$, where $\vec{r}_1$ and
$\vec{r}_2$ are the time dependent positions of the spheres. $F(\dots)$ is the
normal component of the vectorial interaction force $F=\vec{F}\cdot\hat{e}$ with
the unit vector
$\hat{e}=\left(\vec{r}_1-\vec{r}_2\right)/\abs{\vec{r}_1-\vec{r}_2}$. For
non-adhesive viscoelastic spheres it reads \cite{brilliantov1996}
\begin{equation}
\label{eq:viscoelastic}
F=F^\text{el}+F^\text{dis}=\min\left(
0, 
-\rho\xi^{3/2}-\frac{3}{2} A\rho\sqrt { \xi } \dot{ \xi }\right)\,
\end{equation}
with
\begin{equation}
\label{eq:paramEl}
\rho\equiv\frac{2Y\sqrt{R_\text{eff}}}{3(1-\nu^2)}
\end{equation}
and $Y$, $\nu$ and $R_\text{eff}$ stand for the Young modulus, the Poisson
ratio and the effective radius $R_\text{eff}\equiv R_1R_2/(R_1+R_2)$,
respectively. The dissipative constant $A$ is a function of the elastic and
viscous material parameters \cite{brilliantov1996}. The $\min(\dots)$ function
assures that the force is always repulsive.

The elastic part in Eq. \eqref{eq:viscoelastic}, $F^\text{el}$, is the Hertz 
contact force \cite{hertz1882} while its dissipative part, $F^\text{dis}$, was
first motivated in \cite{kuwabara1987} and then rigorously derived in
\cite{brilliantov1996,morgado1997}, where only the approach in
\cite{brilliantov1996} lead to an analytic expression of the material parameter
$A$.

While the knowledge of the interaction force, Eq. \eqref{eq:viscoelastic} is 
sufficient to perform Molecular Dynamics simulations (MD), the coefficient of
restitution is needed to perform much more efficient event-driven MD and Direct
Simulation Monte Carlo (DSMC) as well as for the Kinetic Theory. By disregarding
the dynamics of the collision process and idealizing the collision as an
instantaneous event, the coefficient of restitution relates the postcollisional
normal velocity, $\dot{\xi}^{\prime}$, to the normal component of the
(precollisional) impact velocity, $v$,
\begin{equation}
\label{eq:epsDef}
\varepsilon\equiv-\dot{\xi}^{\prime}/v\,.
\end{equation}

In general, the coefficient of restitution is not a constant but depends on the 
details of the interaction force and the impact velocity. It can be obtained by
integrating Eq. 
\eqref{eq:newton} with the initial conditions $\xi(0)=0$ and $\dot{\xi}(0)=v$, 
assuming that the spheres start contacting at $t=0$. The coefficient of
restitution is then obtained from 
\begin{equation}
\label{eq:force2COR}
\varepsilon=-\dot{\xi}(t_c)/v\,,
\end{equation}
where the duration of the collision, $t_c$, is determined by the condition
\begin{equation}
\label{eq:tEndAdvanced}
\ddot{\xi}(t_c)=0\,\qquad t_c>0\,,
\end{equation}
that is, the collision terminates at time $t_c$ when the interaction force vanishes.

Solving the set of equations (\ref{eq:force2COR},\ref{eq:tEndAdvanced}) is a 
complicated problem which was solved rigorously in \cite{schwager2008}. The
solution reads
\begin{equation}
\label{eq:epsSeriesAdvanced}
\varepsilon=1+\sum_{k=0}^\infty h_k \left(\beta^{1/2}v^{1/10}\right)^k
\equiv 1+\sum_{k=0}^\infty h_k v_\ast^k,
\end{equation}
where we define the shorthand $v_*$ and with
\begin{equation}
\label{eq:betadef}
\beta=\frac{3}{2}A\left(\frac{\rho}{m_\text{eff}}\right)^{2/5}\,.
\end{equation}
This solution is {\em exact} since all coefficients $h_k$ are analytically 
known (see \cite{schwager2008}). It is, moreover, {\em universal} since all
material and particle properties are covered by $\beta$, that is, the $h_k$ are
pure numbers which are independent of the material and particle properties. 

Albeit exact, there are two main problems with the solution,
\eqRef{eq:epsSeriesAdvanced}, which prohibit its application in efficient MD or
DSMC simulations: First, it converges extremely slowly. To obtain $\varepsilon$
up to quadratic order in $v$ we need 20 terms of the series expansion. Second,
wherever we truncate the series at some order $k_c$, 
\eqRef{eq:epsSeriesAdvanced} diverges to $\varepsilon\to\pm\infty$, depending on
the sign of $h_k$.
\begin{figure}[t]
\includegraphics[width=\figWidth]
{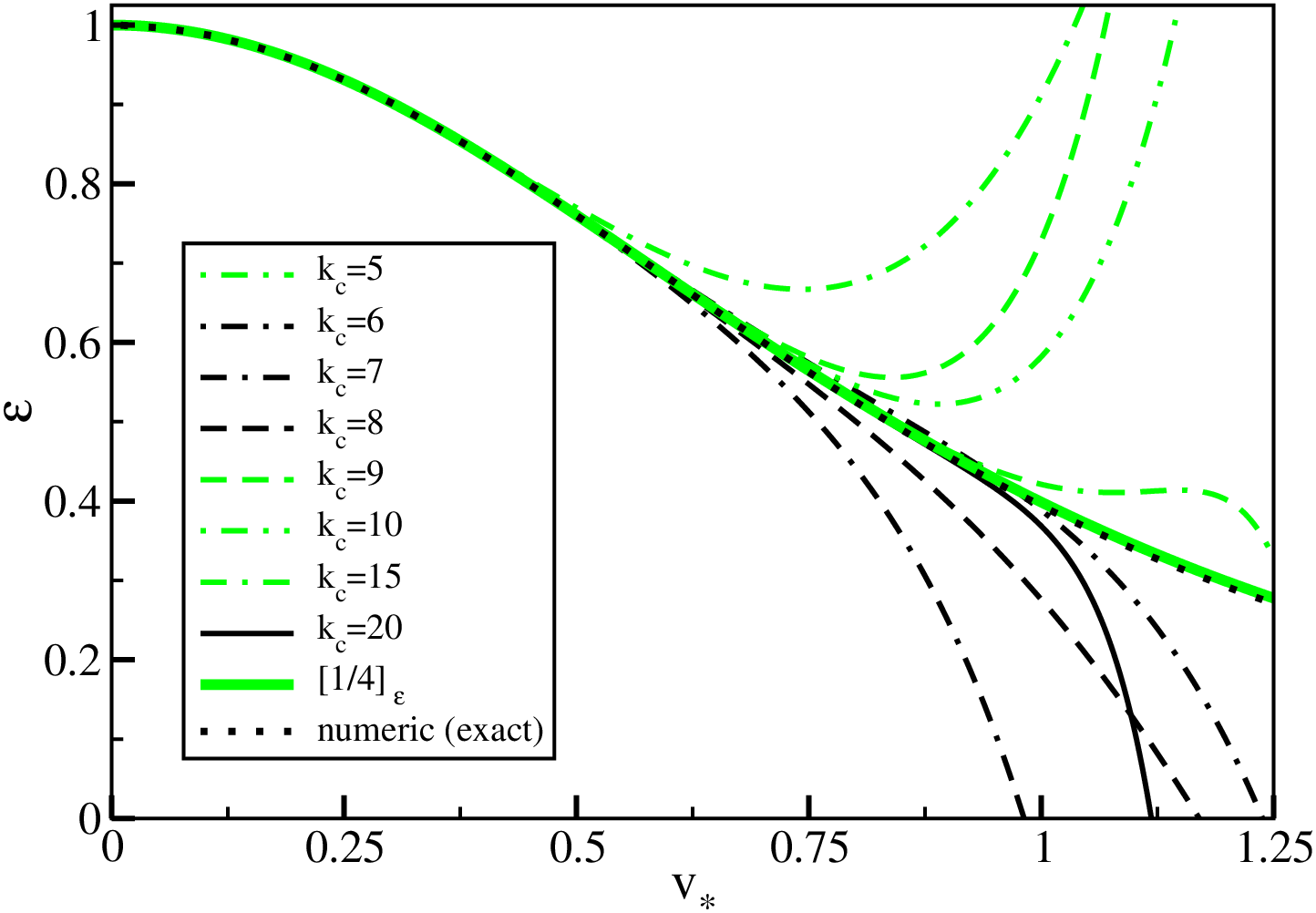}
\caption[fig1]{(color online) Coefficient of restitution, $\varepsilon$, over
$v_\ast\equiv
\beta^{1/2}v^{1/10}$. The analytic solution, Eq. \eqref{eq:epsSeriesAdvanced},
truncated at different order $k_c$ leads to divergence.
The dotted line shows
$\varepsilon$ as it follows from the numerical solution of Eqs.
(\ref{eq:force2COR},\ref{eq:tEndAdvanced}). It almost coincides with the thick
green line showing the Pad\'e approximant $[1/4]_\varepsilon$, Eq.
\eqref{eq:padeDef},
to the analytical solution, Eq. \eqref{eq:epsSeriesAdvanced} (for explanation
see the text below).}
\label{fig:seriesDiverging}
\end{figure}

The divergence of the truncated series is a serious problem: Given the very 
accurate experimental data by Bridges et al. \cite{bridges1984} for the
coefficient of restitution of ice balls at very low temperature whose material
and particle properties correspond to 
$\betaIceThreeSix$.
From Fig. \ref{fig:seriesDiverging} we see that the series truncated at order
$k_c=20$ starts deviating at $v_*\approx 1$ corresponding to the impact velocity
$v=v_*^{10}/\beta^5\approx\vMaxIce$. That is, for typical impact
velocities
of $v\sim 1$ m/sec we would need to go to impractical high truncation order. 
\medskip

From an approximative expression for the coefficient of restitution for 
applications in efficient MD and DSMC simulations, we request that a) the
approximative solution is close to the correct solution, b) it can be computed
efficiently, that is, it contains only a small number of universal coefficients
which are independent of the material and particle properties, and c) the
representation must not reveal divergencies unlike the truncated series, Eq.
\eqref{eq:epsSeriesAdvanced}, shown in Fig. \ref{fig:seriesDiverging}.
\medskip

\begin{figure}[htb]
\includegraphics[width=\figWidth]
{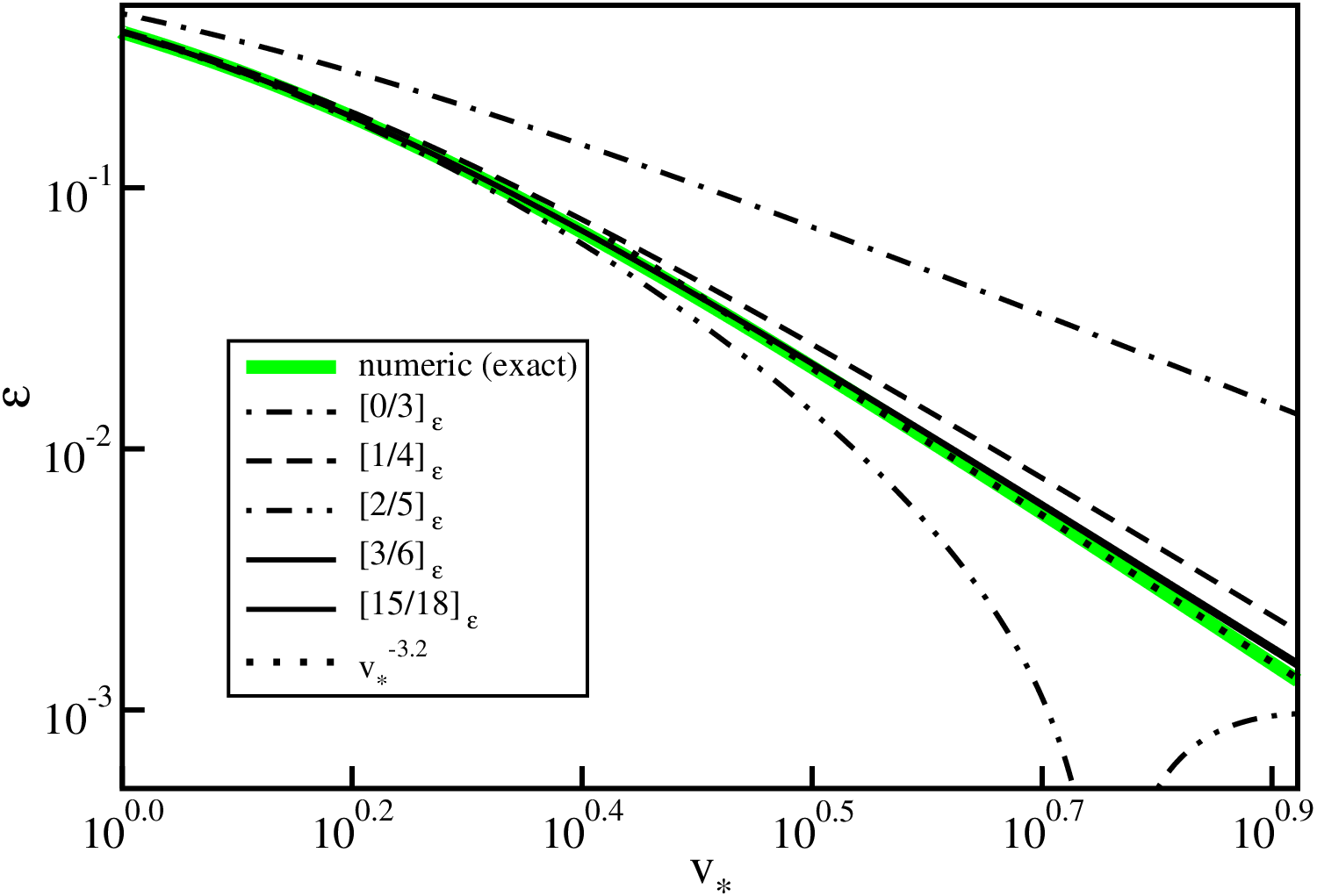}
\caption[fig2]{(color online) Coefficient of restitution $\varepsilon$ for 
large $v_\ast$. The thick green line shows the numerical solution of Eq.
\eqref{eq:equationOfMotionNatUnit} revealing the asymptotic behavior
$\varepsilon = v_\ast^{-3.2}$ (dotted line). Additionally
various Pad\'{e} approximants, Eq. \eqref{eq:padeDef}, of the analytical
solution, Eq. \eqref{eq:epsSeriesAdvanced} are shown (discussion see text
below). The Pad\'e approximants $[3/6]_\varepsilon$ and $[15/18]_\varepsilon$
(virtually identical)
agree almost perfectly with the exact solution.} 
\label{fig:tryingDifferentPade}
\end{figure}

\noindent{\em Numerical solution.} As described in 
\cite{schwager2008,ramirez1999}, \eqRef{eq:newton}
with the interaction force \eqRef{eq:viscoelastic} and the corresponding 
initial conditions may be scaled to 
\begin{equation}
\label{eq:equationOfMotionNatUnit}
\ddot{x}+x^{3/2}+v_\ast^2\,\dot{x}\sqrt{x}=0\,,\quad x(0)=0\,,\quad \dot{x}(0)=1
\end{equation}
with the only free parameter $v_*\equiv \beta^{1/2}v^{1/10}$. Compression and time are scaled by $x\equiv\xi/[(\rho/m_\text{eff})^{-2/5}v^{4/5}]$ and
$\tau\equiv t/[(\rho/m_\text{eff})^{-2/5}v^{-1/5}]$. From the 
numerical solution of \eqRef{eq:equationOfMotionNatUnit} we determine
$\varepsilon(v_\ast)$ via \eqRef{eq:force2COR}:
$\varepsilon=-\dot{\xi}(t_c)/v=-\dot{x}(\tau_c)$, where $\tau_c$ is obtained from the condition $\ddot{x}(\tau_c)=0$,
$\tau_c>0$. Apart from numerical errors, this solution is exact and may serve as
a benchmark for our approximative solution, even for large values of $v_\ast$.
Using the numerical solution we find the asymptotical behavior 
\begin{equation}
  \label{eq:asymp}
\lim_{v_\ast\to \infty}\varepsilon(v_\ast)= v_\ast^{-3.2}  
\end{equation}
for large $v_\ast$, in agreement with \cite{schwager2008}, see 
Fig. \ref{fig:tryingDifferentPade}. 
\medskip

\noindent{\em Pad\'{e} approximants.} Using the analytical solution, 
Eq. \eqref{eq:epsSeriesAdvanced}, and the asymptotics, Eq. \eqref{eq:asymp}, we
construct an approximative expression for $\varepsilon(v)$ which agrees with the
analytical solution for the entire range of definition, $v\in(0,\infty)$, and is
thus much more suitable for numerical simulations. The Pad\'{e} approximant
$[m/n]_\varepsilon(v_*)$ approximates the $m+n$ times differentiable function
$\varepsilon(v_*)$ by a rational function
\begin{equation}
\label{eq:padeDef}
[m/n]_\varepsilon(v_*)=\frac{\sum_{i=0}^{m}a_iv_*^i}{\sum_{i=0}^{n}b_iv_*^i}
\end{equation}
in a way that the Maclaurin series of the approximant and of the approximated 
function match up to order $m+n$: $\varepsilon(0)=[m/n]_{\varepsilon}(0)$,
$\varepsilon^\prime(0)=[m/n]_{\varepsilon}^\prime(0)$, \dots,
$\varepsilon^{(m+n)}(0)=[m/n]_{\varepsilon}^{(m+n)}(0)$. Asymptotically, the
Pad\'{e} approximant behaves like a power law, $\lim_{v_*\to
\infty}[m/n]_\varepsilon\sim v_*^{m-n}$. These properties allow to represent the
function $\varepsilon(v_*)$ similar to a Taylor expansion for small arguments
and asymptotically as a power law, thus, convergent if $m<n$, see Ref. \cite{pade}. 

Since $\varepsilon\sim v_*^\alpha$ with $\alpha\approx-3$ 
(see Eq. \eqref{eq:asymp} and Fig. \ref{fig:tryingDifferentPade}) we chose a
Pad\'e approximation $[m/m+3]_\varepsilon$. To find an accurate yet compact
approximant to Eq. \eqref{eq:epsSeriesAdvanced} we start at $m=0$ and increase
the order until sufficient agreement with the exact solution is achieved. The
result is shown in Fig. \ref{fig:tryingDifferentPade}: $[0/3]_\varepsilon$ is
certainly not acceptable, $[1/4]_\varepsilon$ offers a good tradeoff between
simplicity and accuracy. $[2/5]_\varepsilon$ reveals a pole at $v_*\approx
5.68$, therefore, it is suitable only for small impact velocity, $v_*\lesssim
10^{0.3}$. For ice spheres as described in \cite{bridges1984} this implies
$v\lesssim \vMaxIcePadeTwoFife$. The next order, $[3/6]_\varepsilon$, offers
almost perfect agreement with the benchmark. We checked all orders up to
$[25/28]_\varepsilon$ and could not find any significant improvement as compared
to $[3/6]_\varepsilon$. As an example, $[15/18]_\varepsilon$ is shown in Fig.
\ref{fig:tryingDifferentPade}.

\begin{figure}[htb]
\includegraphics[width=\figWidth]
{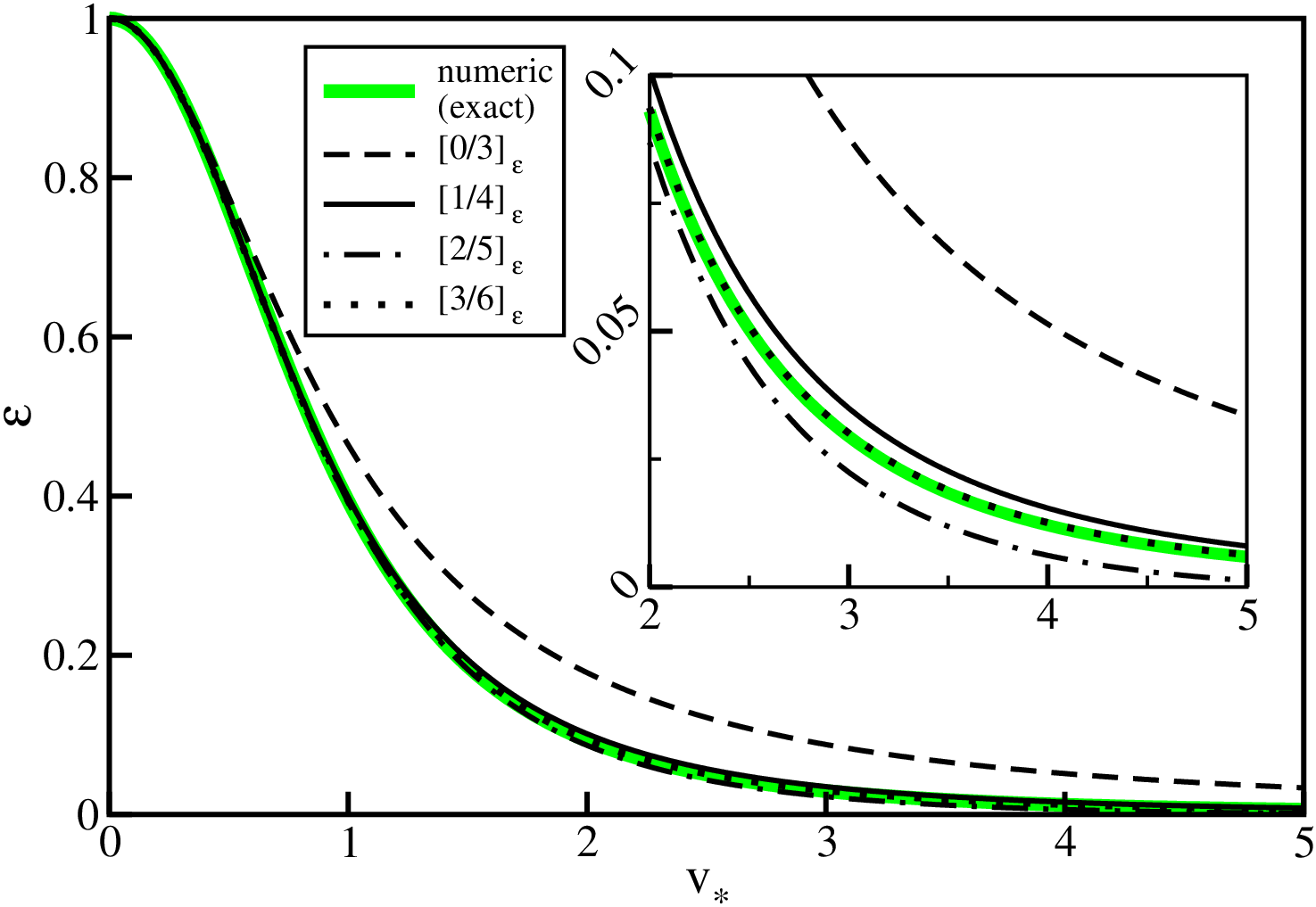}
\caption[fig3]{(color online) Coefficient of restitution $\varepsilon$ over 
$v_\ast$. The first four Pad\'e approximants are shown together with the
numerical (exact) solution. The inset shows a magnification. The order
$[3/6]_\varepsilon$ (dotted line) coincides almost perfectly with the exact
solution in the entire range of definition.}
\label{fig:tryingDifferentPadeLin}
\end{figure}

Table \ref{tab:padeCoeffs} displays the coefficients $a_i$ and $b_i$ for 
the relevant Pad\'e approximants $[m/m+3]_\varepsilon$, $m\in\{0,1,2,3\}$ and 
\begin{table}[htb]
\begin{ruledtabular}
\begin{tabular}{lllll}
\multicolumn{1}{c}{$m$}&\multicolumn{1}{c}{$n$}&\multicolumn{1}{l}{$a_i$}
&\multicolumn{1}{l}{$b_i$}\\ 
\hline$0$&$3$&$a_0=1.$&$b_0=1$ & $b_2=1.15345$\\
& & & $b_1=0$ & $b_3=0$\\
\hline
$1$&$4$&$a_0=1.$&$b_0=1.$ & $b_3=0.577977$ \\
&&$a_1=0.501086$&$b_1=0.501086$ & $b_4=0.532178$\\
&&&$b_2=1.15345$\\
\hline
$2$&$5$&$a_0=1.$&$b_0=1.$ & $b_3=0.638466$ \\
&&$a_1=0.553528$&$b_1=0.553528$ & $b_4=0.384023$\\
&&$a_2=-0.128445$&$b_2=1.025$ & $b_5=0.027908$\\
\hline
$3$&$6$&$a_0=1.$&$b_0=1.$ &$b_4=1.19449$\\
&&$a_1=1.07232$&$b_1=1.07232$ &$b_5=0.467273$\\
&&$a_2=0.574198$&$b_2=1.72765$ &$b_6=0.235585$\\
&&$a_3=0.141552$&$b_3=1.37842$\\ 
\end{tabular}
\end{ruledtabular}
\caption[padeCoeffs]{Coefficients of the Pad\'{e} approximants 
$[m/m+3]_\varepsilon$ for $m\in\{0,1,2,3\}$. $[2/5]_\varepsilon$ reveals a pole
at
$v_*\approx 5.6801$.}
\label{tab:padeCoeffs}
\end{table}
Fig. \ref{fig:tryingDifferentPadeLin} shows these Pad\'{e} approximants 
together with the exact (numerical) solution. Again $[1/4]_\varepsilon$ and
$[3/6]_\varepsilon$ turn out to be good compromizes between accuracy and
simplicity.

\medskip

\noindent{\em Conclusion.} The universal exact solution, 
Eq. \eqref{eq:epsSeriesAdvanced}, for the coefficient of restitution of smooth
viscoelastic spheres cannot be applied directly in eMD and DSMC simulations
since
the series diverges for any {\em finite} truncation order. We have shown that
the Pad\'e approximations of order $[1/4]_\varepsilon$ and $[3/6]_\varepsilon$
are suitable to represent the coefficient of restitution over the entire range
of impact velocities including its asymptotic behavior up to an excellent
accuracy and we provided the constants of this approximation. Similar as the
full solution, Eq. \eqref{eq:epsSeriesAdvanced}, the Pad\'e expansion is
universal, that is, the constants $a_i$ and $b_i$ are universal. They neither
depend on material properties (Young modulus, Poisson ratio, dissipative
constant) nor on particle properties (radii, masses). All non-universal
parameters enter exclusively via $\beta$, Eq. \eqref{eq:betadef}, which in turn
enters the argument of the Pad\'e expansion via $v_*=\beta^{1/2}v^{1/10}$ with
$v$ being the impact velocity in physical units ($cm/sec$). Thus, the presented
Pad\'e
approximation can be conveniently applied in numerical simulations.

\begin{figure}[htb]
\includegraphics[width=\figWidth]
{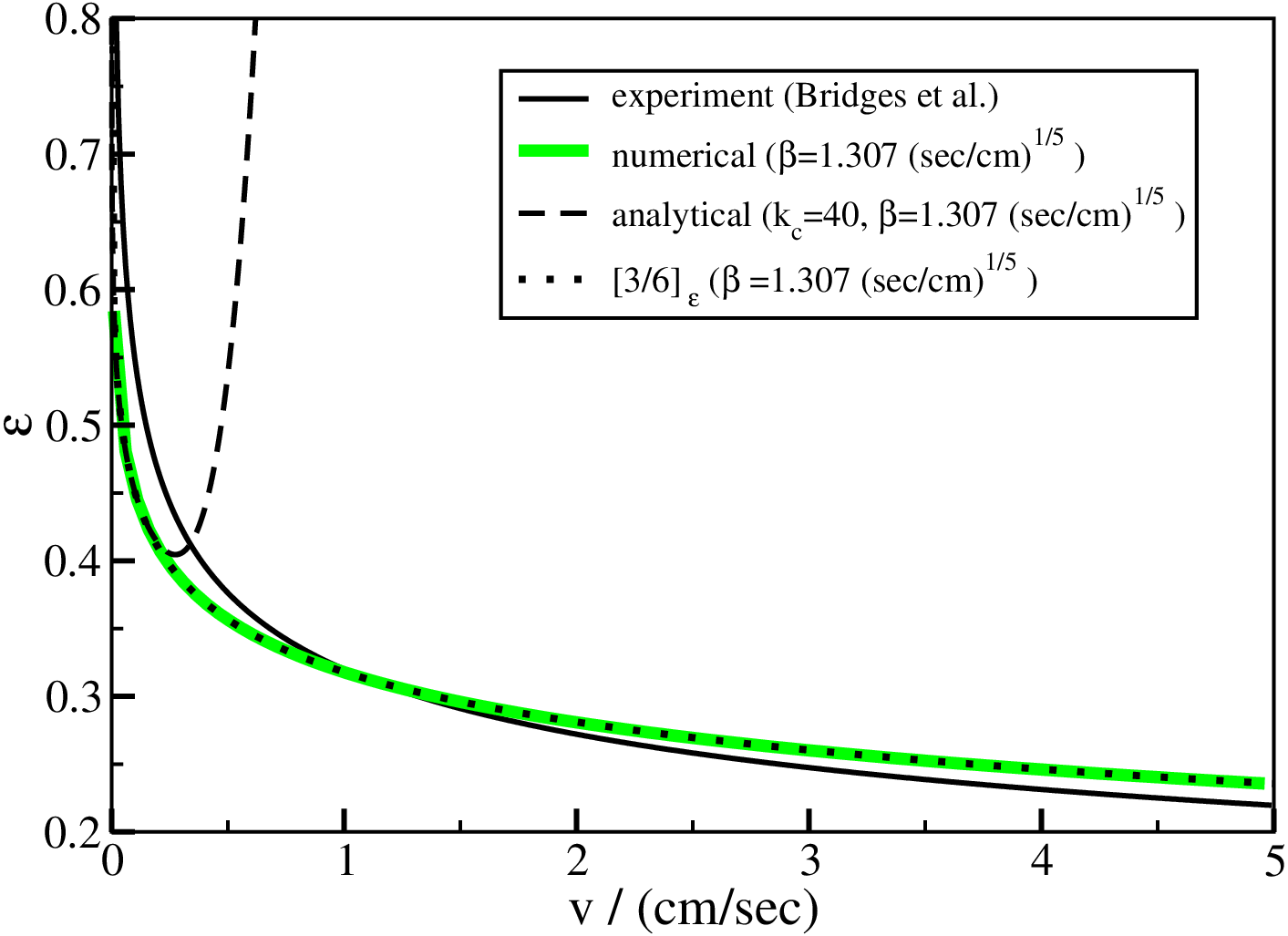}
\caption[fig4]{(color online) Coefficient of restitution $\varepsilon$ as a
function of the impact velocity $v$. The Pad\'e approximant $[3/6]_\varepsilon$ 
(dotted line) agrees almost perfectly with the numerical integration of Newton's
equation, Eqs. (\ref{eq:newton}-\ref{eq:tEndAdvanced}) in the entire range of
impact velocity, $v$ (physical units), while the analytical solution, Eq.
\eqref{eq:epsSeriesAdvanced}, truncated at order as large as $k_c=40$ diverges
at
$v\approx 0.3 $\,cm/sec. For the material constant, $\betaIceThreeSix$, we
used
the experimental values by Bridges et al. \cite{bridges1984} for the collision
of ice spheres at low temperature.}
\label{fig:expVsAnaVsNum}
\end{figure}
The precision of the approximant can be assessed in 
Fig. \ref{fig:expVsAnaVsNum} which shows the Pad\'e approximation together with
the numerical integration of Newton's equation, Eq. \eqref{eq:newton} in
combination with Eqs. (\ref{eq:viscoelastic}-\ref{eq:tEndAdvanced}), and with
the divergent analytical solution, Eq. \eqref{eq:epsSeriesAdvanced}, truncated
at order as large as $k_c=40$.  We see that over the entire range of definition,
the Pad\'e approximation coincides almost perfectly with the numerical solution
and with the truncated analytical solution up to $v\approx 0.3 $\,cm/sec where
it
starts to diverge. For the material constant, $\betaIceThreeSix$, we used the
experimental values by Bridges et al. \cite{bridges1984} for the collision of
ice spheres at low temperature. The corresponding data is also shown in the
plot. While the agreement between the exact analytical result, the numerical
integration and the Pad\'e approximant is remarkable, the experimental data
slightly deviates. This deviation is not surprising since besides
viscoelasticity, described by the force Eq. \eqref{eq:viscoelastic}, other
forces may contribute, such as surface forces, plastic deformation, adhesion
etc. 

\noindent{\em Acknowledgement.} The authors gratefully acknowledge the support 
of the Cluster of Excellence 'Engineering of Advanced Materials' at the 
University of Erlangen-Nuremberg, which is funded by the German Research 
Foundation (DFG) within the framework of its 'Excellence Initiative'.

%

\end{document}